\documentclass{amsart}

\theoremstyle{plain}
\newtheorem{theorem}{Theorem}[section]

\theoremstyle{definition}
\newtheorem*{definition}{Definition}

\theoremstyle{remark}
\newtheorem{remark}{Remark}

\numberwithin{equation}{section}

\begin{document}

\title{Nonlinear stability analysis of the Emden-Fowler equation}

\author{C. G. B\"OHMER}
\address{Department of Mathematics, University College London,\\
  Gower Street, London, WC1E 6BT, United Kingdom}
\email{c.boehmer@ucl.ac.uk}

\author{T. HARKO}
\address{Department of Physics and Center for Theoretical\\
  and Computational Physics, The University of Hong Kong,\\
  Pok Fu Lam Road, Hong Kong, Hong Kong SAR, P. R. China}
\email{harko@hkucc.hku.hk}



\date{14 April 2010}

\begin{abstract}
In this paper we qualitatively study radial solutions of the semilinear elliptic equation $\Delta u + u^n = 0$ with $u(0)=1$ and $u'(0)=0$ on the positive real line, called the Emden-Fowler or Lane-Emden equation. This equation is of great importance in Newtonian astrophysics and the constant $n$ is called the polytropic index.

By introducing a set of new variables, the Emden-Fowler equation can be written as an autonomous system of two ordinary differential equations which can be analyzed using linear and nonlinear stability analysis. We perform the study of stability by using linear stability analysis, the Jacobi stability analysis (Kosambi-Cartan-Chern theory) and the Lyapunov function method. Depending on the values of $n$ these different methods yield different results. We identify a parameter range for $n$ where all three methods imply stability.
\end{abstract}

\maketitle

\section{Introduction}

The study of equations of the form $\Delta u + f(u) = 0$ in bounded or unbounded domain has received great attention over the last decades. Many natural processes are described by equations of this type. Of particular interest for astrophysics are the radial solutions of the Emden-Fowler equation where $f(u)=u^n$ and the basic equation becomes a second order non-linear ordinary differential equation. The mathematical properties have been extensively studied in the past, see for instance~\cite{Chandrasekhar:1939,Wong:1975,Collins:1977,Kimura:1981aa,Kimura:1981bb,Kimura:1981cc,Horedt:1986aa,Horedt:1986bb,Horedt:1987aa,Horedt:1987bb,Horedt:1987cc,Blaga:1996aa,Blaga:1996bb,Kuzin:1997,Schaudt:2000,Heinzle:2002sk,Benguria:2003,Horedt:2004,Blaga:2005,Govinder:2007}. In particular, the qualitative phase space analysis of the dynamical system associated to the Emden-Fowler equation proved to be very useful for the understanding of the general mathematical and physical properties of its solutions.

It is the purpose of the present paper to consider the qualitative study of the stability properties of the Emden-Fowler equation which can be transformed to an autonomous system of differential equations. After using linear perturbation theory to analyze the stability of the critical points, we investigate these equations in geometric terms by using \textit{the general path-space theory of Kosambi-Cartan-Chern (KCC-theory)}, inspired by the geometry of a Finsler space~\cite{Kosambi:1933,Cartan:1933,Chern:1939}. The key idea behind this approach is to geometrize the differential equation itself, rather than analyzing its solutions directly. This yields to a manifold (Finsler space) similar to phase-space, whose curvature properties determine the behavior of \textit{all} solutions in a non-linear setting. The KCC theory has been applied for the study of different physical, biochemical or technical systems (see~\cite{Antonelli:1993,Antonelli:2000,Antonelli:2001aa,Antonelli:2001bb,Antonelli:2003,Sabau:2005aa,Sabau:2005bb,Yajima:2007,Harko:2008ak} and references therein). Moreover, to complete our study, we also use the Lyapunov function method to analyze the stability properties. By combining these three methods, we identify a previously unrecognized parameter range for $n$ where stability is implied by all methods.

The paper is organized as follows. The basic equations and the linear perturbation theory are presented in Section~\ref{sec:basic}. We review the mathematical formalism of the KCC theory and apply it in Section~\ref{kcc}. The Lyapunov function method is used in Section~\ref{sec:lya}. We discuss and conclude our results in Section~\ref{sec:dis}.

\section{Basic equations}
\label{sec:basic}

\subsection{Emden-Fowler equation}

The properties of the static Newtonian astrophysical objects can be completely described by the gravitational structure equations, which are the mass continuity equation and the equation of the hydrostatic equilibrium.
\begin{definition}[Gravitational structure equations of Newtonian stars]The basic equations describing the equilibrium properties of Newtonian stars are given by
\begin{align}
  \frac{dm(r)}{dr} &= 4 \pi \rho(r) r^{2},
  \label{5}\\
  \frac{dp(r)}{dr} &= - \frac{G m(r)}{r^{2}}\rho(r) ,
  \label{6}
\end{align}
where $\rho (r)\geq 0$ is the density, $p(r)\geq 0$ is the thermodynamical pressure and $m(r)$ is the mass inside radius $r$, respectively, satisfying the condition $m(r)\geq 0,\forall r\geq 0$.
\end{definition}
The system of differential equations, given by Eqs.~(\ref{5}) and (\ref{6}), contains three unknown functions, but it consists of only two equations. To close it one must prescribe an equation of state, $p=p(\rho)$, which relates the thermodynamical pressure to the density of the fluid. We assume that the equation of state is sufficiently smooth for all $p>0$.

\begin{remark}
From an astrophysical point of view it seems also well motivated to prescribe in some situations the density distribution of matter $\rho(r)$, rather than the equation of state. However, the general treatment of the problem shows that this approach may yield solutions with a non-regular center, which are therefore un-physical.
\end{remark}

\begin{definition}[Polytropic perfect fluid]
An isotropic fluid distribution for which the pressure and the density are related by a power law of the form
\begin{align}
  p=K\rho ^{1+1/n},
\end{align}
is called a polytropic perfect fluid. $K$ and $n$ are constants, and $n \neq 0$ is called the polytropic index.
\end{definition}

\begin{theorem}[Emden-Fowler equation]
For a polytropic perfect fluid sphere the gravitational structure equations~(\ref{5}) and (\ref{6}) are equivalent to the following single second order non-linear differential equation in $\theta(\xi)$, called the Emden-Fowler equation
\begin{align}
  \theta'' + \frac{2}{\xi} \theta' + \theta^{n} = 0,
  \label{LE}
\end{align}
where the variable $\theta $ is related to the density by means of the definition $\rho = \rho _{c} \theta^{n}$, and $n$ is called the polytropic index. The initial conditions for the Emden-Fowler equation are $\theta(0)=1$ and $\theta'(0)=0$, respectively, where the prime denotes the derivative with respect to the dimensionless radial coordinate $\xi $.
\end{theorem}

\begin{proof}
By eliminating the mass between the structure equations~(\ref{5}) and (\ref{6}) we obtain a single second order non-linear differential equation
\begin{align}
  \frac{1}{r^{2}}\frac{d}{dr}\left(\frac{r^{2}}{\rho }\frac{dp}{dr}\right)
  =-4\pi G\rho.
  \label{7}
\end{align}

It is convenient to represent the density in terms of a new dimensionless variable $\theta$ by
\begin{align}
  \rho = \rho _{c} \theta^{n},
\end{align}
giving for the pressure $p=K\rho _{c}^{1+1/n}\theta ^{n+1}$. Let us furthermore introduce a new dimensionless radial coordinate $\xi $ such that
\begin{align}
  r=\alpha \xi, \qquad
  \alpha =\sqrt{\frac{(n+1)K\rho_{c}^{1/n-1}}{4\pi G}}, \qquad
  n\neq -1.
\end{align}
In these new variables Eq.~(\ref{7}) takes the form of the Emden-Fowler equation of index $n$.
\end{proof}

\begin{remark}
In the limiting case $n \rightarrow 0$, the Emden-Fowler equation is solved by $\left.\theta(\xi)\right|_{n=0}=1-\xi^2/6$. For $n=1$, the Emden-Fowler equation~(\ref{LE}) becomes linear and is solved by $\left.\theta(\xi)\right|_{n=1}=\sin(\xi)/\xi$. The only known solution to the non-linear equation is when $n=5$, and it is given by $\left.\theta(\xi)\right|_{n=5}=1/\sqrt{1+\xi^2/3}$.
\end{remark}

\subsection{The Emden-Fowler equation as an autonomous system}

In order to study the stability of the equilibrium points of the Emden-Fowler, we rewrite it in the form of an autonomous system of differential equations.

\begin{theorem}
The Emden-Fowler equation~(\ref{LE}) is equivalent to the following autonomous system of differential equations
\begin{align}
  \frac{dw}{dt} &= q,
  \label{eqw}
  \end{align}
  \begin{align}
  \frac{dq}{dt} &= -2 G^{1}(w,q),
  \label{syst}
\end{align}
with the function $G^1(w,g)$ given by
\begin{align}
  G^{1}(w,q) =\frac{1}{2}\left[ -\frac{n-5}{n-1}q+
  \frac{2(3-n)}{(n-1)^{2}}w+B^{n-1}w^{n}\right].
  \label{eqn:g1}
\end{align}
\end{theorem}

\begin{proof}
Let us start by introducing a set of new variables $(w,t)$ defined as
\begin{align}
  \theta(\xi) = B\xi^{2/(1-n)}w(\xi),\qquad \xi = \xi_s\,e^{-t},
\end{align}
where $\xi _S$ is the value of $\xi $ at the star's surface, and $B>0$ is a constant (which generally  can be set to one without any loss of generality), the Emden-Fowler equation is equivalent with the following second order differential equation,
\begin{align}
  \frac{d^{2}w}{dt^{2}}-\frac{n-5}{n-1}\frac{dw}{dt}-
  \frac{2(n-3)}{(n-1)^{2}}w+B^{n-1}w^{n}=0,\qquad
  n \neq 1.
  \label{u}
\end{align}
In the following we will restrict the range of the polytropic index $n$ to the range $n > 1$. From a mathematical point of view, Eq.~(\ref{u}) is a second order non-linear differential equation of the form $d^{2}w/dt^{2}+2G^{1}(w,dw/dt)=0$, with $G^1$ given by~(\ref{eqn:g1}). Note that $G^1(0,0)=0$. By introducing a new variable $q=dw/dt$, this second order differential equation is equivalent to the  first order autonomous system of equations given by Eqs.~(\ref{eqw}) and (\ref{syst}).
\end{proof}
\begin{remark}
For the initial Emden-Fowler equation the range of the radial dimensionless variable $\xi $ is $\xi \in [0,\xi_s]$. The range of the new variable $t=\ln(\xi_s/\xi)$ is $t \in (\infty,0]$, so that $t=0$ at the surface of the star, where $\xi =\xi_s$, and $t \rightarrow  \infty$ at the center of the star ($\xi=0$). Note that if the solution describes an object of infinite extend ($n \geq 5$), then $\xi \in [0,\infty)$ and $t\in (\infty,-\infty)$.

At the surface of the star (the vanishing pressure surface) the function $w(t)$ takes the value $w(t=0)=0$ since $\theta(\xi_s)=0$, while $w'(t=0)$ is given by $w'(t=0)=\theta'(\xi_s)\xi_s^{2/(n-1)}/B$. At the center of the star we have $\lim_{t\rightarrow \infty }w\left( t\right) =0$ and $\lim_{t\rightarrow\infty }w^{\prime }(t)=0$, respectively.
\end{remark}

The critical points of this dynamical system all reside on the $q=0$ line. They are given by the solutions of the equation
\begin{align}
  G^{1}(w,0) = \frac{1}{2}
  \left[-\frac{2(n-3)}{(n-1)^{2}}w+B^{n-1}w^{n}\right] = 0.
  \label{eqal}
\end{align}
Therefore for the critical points $X_{i}=(w_{0},q_{0})$ of the system~(\ref{eqw}) and (\ref{syst}) we firstly find the point
\begin{align}
  X_{0}=(0,0),
  \label{critp1}
\end{align}
whose value is $n$ independent.

For $w \neq 0$ Eq.~(\ref{eqal}) is equivalent to
\begin{align}
  \frac{2(n-3)}{(n-1)^{2}} = B^{n-1}w^{n-1}.
  \label{eqn:g2}
\end{align}
Hence, for $1 < n < 3$ we find
\begin{align}
  X_{n} &= \left(\sqrt[n-1]{-1} \frac{1}{B}\sqrt[n-1]{\frac{2(3-n)}{(n-1)^2}},0\right).
  \label{critp4}
\end{align}
while for $n>3$ we have
\begin{align}
  X_{n}=\left(\frac{1}{B}\left[\frac{2\left(n-3\right)}{(n-1)^{2}}\right]^{1/(n-1)},0\right).
  \label{critp3}
\end{align}

\subsection{Linear stability analysis of the equilibrium points}

To characterize the nature of the critical points~(\ref{critp1})--(\ref{critp3}) we use, as a first method, the linear stability analysis~\cite{Boyce:1992}. The results of this analysis can be summarized in the following theorem.

\begin{theorem}
Let the dynamical system~(\ref{eqw}) and (\ref{syst}), with critical points~(\ref{critp1}) and~(\ref{critp3}), be given. Then the point $X_0$ is stable if $1 < n < 3$ and unstable otherwise. The point $X_n$ is stable for $3 < n < 5$ and unstable otherwise.
\end{theorem}

\begin{proof}
The eigenvalues of the Jacobian matrix of the first derivatives of the dynamical system,
\begin{align}
  \begin{pmatrix} 0 & 1 \\
  2(n-3)/(n-1)^2 - n B^{n-1}w^{n-1} & \,(n-5)/(n-1)
  \end{pmatrix},
\end{align}
are given by
\begin{align}
  \lambda_{\pm} = \frac{n-5}{2(n-1)} \pm \frac{1}{2}
  \sqrt{1-4n B^{n-1}w^{n-1}}.
\end{align}
At the point $X_0$ we therefore find
\begin{align}
  \lambda_{\pm} = \frac{1}{2}\left(\frac{n-5}{n-1} \pm 1\right),
\end{align}
which gives the following for the eigenvalues
\begin{alignat}{3}
  &\lambda_{+} < 0, &\qquad  &\lambda_{-} < 0,
  &\qquad 1 &< n < 3,\\
  &\lambda_{+} > 0, &\qquad  &\lambda_{-} < 0,
  &\qquad 3 &< n.
\end{alignat}
These eigenvalues characterize the critical point $X_0$, and we have
\begin{alignat*}{2}
  1 &< n < 3 &\qquad &\text{nodal sink (stable)}\\
  3 &< n &\qquad &\text{saddle point (unstable)}.
\end{alignat*}

At $X_n$ (here we refer to all $n>1$) one finds
\begin{align}
  \lambda_{\pm} = \frac{1}{2(n-1)}(n-5 \pm \sqrt{1+22n-7n^2}).
\end{align}
We obtain the following for the eigenvalues
\begin{alignat}{3}
  &\lambda_{+} > 0, &\qquad  &\lambda_{-} < 0,
  &\qquad 1 &< n < 3,\\
  &\lambda_{+} < 0, &\qquad  &\lambda_{-} < 0,
  &\qquad 3 &< n < (11+8\sqrt{2})/7,\\
  \mathrm{Re}\,&\lambda_{+} < 0, &\qquad  \mathrm{Re}\,&\lambda_{-} < 0,
  &\qquad  (11+8\sqrt{2})/7 &< n < 5,\\
  \mathrm{Re}\,&\lambda_{+} > 0, &\qquad  \mathrm{Re}\,&\lambda_{-} > 0,
  &\qquad  5 &< n.
\end{alignat}
These eigenvalues characterize the critical point $X_n$, and we find
\begin{alignat*}{2}
  1 &< n < 3 &\qquad &\text{saddle point (unstable)}\\
  3 &< n < (11+8\sqrt{2})/7 &\qquad &\text{nodal sink (stable)}\\
  (11+8\sqrt{2})/7 &< n < 5 &\qquad &\text{spiral sink (stable)}\\
  5 &< n &\qquad &\text{spiral source (unstable)}.
\end{alignat*}
This completes the linear stability analysis.
\end{proof}

\section{Kosambi-Cartan-Chern theory and the Jacobi stability}
\label{kcc}

\subsection{Theory}

We recall the basics of Kosambi-Cartan-Chern (KCC) theory to be used in the sequel~\cite{Antonelli:1993,Antonelli:2000,Miron:2001,Antonelli:2001aa,Antonelli:2001bb,Antonelli:2003,Miron:2005,Sabau:2005aa,Sabau:2005bb}. Let $\mathcal{M}$ be a real, smooth $n$-dimensional manifold and let $T\mathcal{M}$ be its tangent bundle. Let us choose $(x^{i}) =(x^{1},x^{2},\ldots,x^{n})$, $(y^{i})=(y^{1},y^{2},\ldots,y^{n})$ and the time $t$ be a $2n+1$ coordinates system of an open connected subset $\Omega $ of the Euclidian $(2n+1)$ dimensional space $\mathbb{R}^{n}\times \mathbb{R}^{n}\times \mathbb{R}^{1}$, where
\begin{align}
  y^{i}=\left( \frac{dx^{1}}{dt},\frac{dx^{2}}{dt},\ldots,\frac{dx^{n}}{dt}
  \right).
\end{align}
We assume that $t$ is an absolute invariant, and therefore the only admissible change of coordinates will be
\begin{align}
  \tilde{t}=t,
  \qquad\tilde{x}^{i}=\tilde{x}^{i}( x^{1},x^{2},\ldots,x^{n}),
  \qquad i \in \{1 ,2,\ldots,n\}.
  \label{ct}
\end{align}

The equations of motion of a dynamical system can be derived from a Lagrangian $L$ via the Euler-Lagrange equations. For a regular Lagrangian the Euler-Lagrange equations are equivalent to a system of second-order differential equations
\begin{align}
  \frac{d^{2}x^{i}}{dt^{2}} + 2G^{i}(x^{j},y^{j},t) = 0,
  \qquad i \in \{1,2,\ldots,n\},
  \label{EM}
\end{align}
where each function $G^{i}(x^{j},y^{j},t)$ is $C^{\infty}$ in a neighborhood of some initial conditions $( x_{0}, y_{0}, t_{0})$ in $\Omega$. The system of Eqs.~(\ref{EM}) is equivalent to a vector field (semi-spray) $S$, where
\begin{align}
  S = y^{i}\frac{\partial}{\partial x^{i}}-2G^{i}(x^{j},y^{j},t)
  \frac{\partial }{\partial y^{i}},
\end{align}
which determines a non-linear connection $N_{j}^{i}$, defined as
\begin{align}
  N_{j}^{i}=\frac{\partial G^{i}}{\partial y^{j}}.
\end{align}
More generally, one can start from an arbitrary system of second-order differential equations of the form~(\ref{EM}), with no \textit{a priori} given Lagrangian, and study the behavior of its trajectories, by analogy with the trajectories of the Euler-Lagrange equations.

If the coordinate transformations given by Eq.~(\ref{ct}) is nonsingular, then the KCC-covariant differential of a vector field $\xi^{i}(x)$ on the open subset $\Omega \subseteq \mathbb{R}^{n}\times \mathbb{R}^{n}\times \mathbb{R}^{1}$ is defined as
\begin{align}
  \frac{D\xi^{i}}{dt} = \frac{d\xi^{i}}{dt}+N_{j}^{i} \xi^{j}.
  \label{KCC}
\end{align}
For $\xi^{i}=y^{i}$ we obtain
\begin{align}
  \frac{Dy^{i}}{dt}=N_{j}^{j}y^{j}-2G^{i}=-\epsilon^{i}.
  \qquad i\in \{1,2,\ldots,n\}
\end{align}
The contravariant vector field $\epsilon^{i}$ on $\Omega$ is called the first KCC invariant.

Let us now vary the trajectories $x^{i}(t)$ of the system~(\ref{EM}) into nearby ones, according to
\begin{align}
  \tilde{x}^{i}(t) = x^{i}(t) + \eta\, \xi^{i}(t),
  \label{var}
\end{align}
where $|\eta| $ is a small parameter, and $\xi^{i}(t)$ are the components of some contravariant vector fields, defined along the path $x^{i}(t)$. Substituting Eqs.~(\ref{var}) into Eqs.~(\ref{EM}), and taking the limit $\eta \rightarrow 0$, we obtain the variational equations
\begin{align}
  \frac{d^{2}\xi^{i}}{dt^{2}} + 2N_{j}^{i}\frac{d\xi ^{j}}{dt} +
  2\frac{\partial G^{i}}{\partial x^{j}}\xi^{j} = 0.
  \label{def}
\end{align}

The covariant form of Eq.~(\ref{def}) is
\begin{align}
  \frac{D^{2}\xi^{i}}{dt^{2}} = P_{j}^{i} \xi^{j},
  \label{JE}
\end{align}
where
\begin{align}
  P_{j}^{i} = -2\frac{\partial G^{i}}{\partial x^{j}} - 2G^{l}G_{jl}^{i} +
  y^{l} \frac{\partial N_{j}^{i}}{\partial x^{l}} + N_{l}^{i}N_{j}^{l} +
  \frac{\partial N_{j}^{i}}{\partial t},
\end{align}
and
\begin{align}
  G_{jl}^{i} = \frac{\partial N_{j}^{i}}{\partial y^{l}}.
\end{align}
Eq.~(\ref{JE}) is the Jacobi equation, $G_{jl}^{i}$ is called the Berwald connection, and $P_{j}^{i}$ is the second KCC-invariant, or the deviation curvature tensor~\cite{Antonelli:1993,Miron:2001,Sabau:2005aa,Sabau:2005bb}. When the system~(\ref{EM}) describes the geodesic equations in either Riemann or Finsler geometry, Eq.~(\ref{JE}) is the Jacobi field equation.

\begin{remark}
One can also define higher order invariants of the system~(\ref{EM}). The third, fourth and fifth invariants are given by
\begin{align}
  P_{jk}^{i} = \frac{1}{3}\left( \frac{\partial P_{j}^{i}}{\partial y^{k}}-
  \frac{\partial P_{k}^{i}}{\partial y^{j}}\right),
  \qquad P_{jkl}^{i} = \frac{\partial P_{jk}^{i}}{\partial y^{l}},
  \qquad D_{jkl}^{i} = \frac{\partial G_{jk}^{i}}{\partial y^{l}}.
\end{align}
The third invariant is interpreted as a torsion tensor, while the fourth and fifth invariants are the Riemann-Christoffel curvature tensor, and the Douglas tensor, respectively~\cite{Antonelli:2000}. In a Berwald space these tensors always exist, and they describe the geometrical properties of a system of second-order differential equations.
\end{remark}

The Jacobi stability for a dynamical system is defined as follows~\cite{Antonelli:2000,Sabau:2005aa,Sabau:2005bb}:

\begin{definition}[Jacobi stability]
If the system of differential equations~(\ref{EM}) satisfies the initial conditions $\|x^{i}(t_{0})-\tilde{x}^{i}(t_{0})\| = 0$, $\|\dot{x}^{i}(t_{0})-\tilde{x}^{i}(t_{0})\| \neq 0$, with respect to the norm $\| \cdot \|$ induced by a positive definite inner product, then the trajectories of~(\ref{EM}) are Jacobi stable if and only if the real parts of the eigenvalues of the deviation tensor $P_{j}^{i}$ are strictly negative everywhere, and Jacobi unstable, otherwise.
\end{definition}

\subsection{Application to the Emden-Fowler equation}

\begin{theorem}
Let the dynamical system~(\ref{eqw}) and (\ref{syst}), with critical points~(\ref{critp1}) and~(\ref{critp3}), be given. Then the point $X_0$ is Jacobi unstable. However, if $n > (11+8\sqrt{2})/7$, the point $X_n$ is Jacobi stable, and unstable otherwise.
\end{theorem}

\begin{proof}
By denoting $x^{1}:=w$ and $y^{1}:=dx^{1}/dt=dw/dt$, Eq.~(\ref{u}) can be written as
\begin{align}
  \frac{d^{2}x^{1}}{dt^{2}}+2G^{1}(x^{1},y^{1}) = 0,
  \label{jac2}
\end{align}
where
\begin{align}
  G^{1}(x^{1},y^{1}) = \frac{1}{2}\left[ -\frac{n-5}{n-1}y^{1}-\frac{2(n-3)}{(n-1)^{2}}x^{1}+B^{n-1}(x^{1})^{n}\right],
  \label{G1}
\end{align}
respectively, which can now be studied by means of the KCC theory. As a first step in the KCC stability analysis of the Newtonian polytropes we obtain the nonlinear connections $N_{1}^{1}=\partial G^{1}/\partial y^{1}$, associated to Eqs.~(\ref{jac2}), and which is given by
\begin{align}
  N_{1}^{1}(x^{1},y^{1}) = -\frac{1}{2}\frac{n-5}{n-1}.
\end{align}
For the Emden-Fowler equation the associated Berwald connection identically vanishes
\begin{align}
  G_{11}^{1} = \frac{\partial N_{1}^{1}}{\partial y^{1}} = 0.
\end{align}

Finally, the second KCC invariant, or the deviation curvature tensor $P_{1}^{1}$, defined as
\begin{align}
  P_{1}^{1}=-2\frac{\partial G^{1}}{\partial x^{1}}-2G^{1}G_{11}^{1}+y^{1}
  \frac{\partial N_{1}^{1}}{\partial x^{1}}+N_{1}^{1}N_{1}^{1}+
  \frac{\partial N_{1}^{1}}{\partial t},
\end{align}
is given by
\begin{align}
  P_{1}^{1}(x^{1},y^{1}) = \frac{1}{4}-nB^{n-1}(x^{1})^{n-1}.
\end{align}
In the initial variables $P_{1}^{1}$ is given by
\begin{align}
  P_{1}^{1}(w,q) = \frac{1}{4} - n B^{n-1} w^{n-1}.
\end{align}

Evaluating $P_{1}^{1}(w,q) $ at the critical points $X_{n}$, given by Eqs.~(\ref{critp1})-(\ref{critp3}), we obtain
\begin{alignat}{2}
  P_{1}^{1}(X_{0}) &= \frac{1}{4}>0,
  &\quad &\forall\,n,\\
  P_{1}^{1}(X_{n}) &= \frac{-7n^{2}+22n+1}{4(n-1)^{2}},
  &\quad &n > 1.
\end{alignat}

Therefore, according to the Jacobi stability theorem, the point $X_0$ is unstable since $P_1^1 > 0$. However, for $n > (11+8\sqrt{2})/7$ the function $P_1^1$ becomes strictly negative and Jacobi stability is established.
\end{proof}

\subsection{Physical interpretation of Jacobi stability}

In the initial variables the deviation curvature tensor $P_{1}^{1}$ is given by
\begin{align}
  P_{1}^{1}(\xi,\theta) = \frac{1}{4} - n\xi^{2}\theta^{n-1}.
\end{align}
$P_{1}^{1}$ can be expressed in a simple form with the use of Milne's homological variables $(u,v)$, see~\cite{Horedt:1987cc,Horedt:2004}, defined as
\begin{align}
  u=-\frac{\xi \theta^{n}}{\theta'},
  \qquad v=-\frac{\xi \theta'}{\theta},
\end{align}
with the use of whom we obtain
\begin{align}
  P_{1}^{1}(u,v) = \frac{1}{4} - n u v.
\end{align}

From a physical point of view $u(r)$, defined as $u(r)=d\ln m(r)/d\ln r=3\rho (r)/\bar{\rho}(r)$, is equal to three times the density of the star at point $r$, divided by the mean density of matter, $\bar{\rho}(r)$, contained in the sphere of radius $r$. The variable $v(r)$, defined as $v(r)=-d\ln p(r)/d\ln r=(3/2)[Gm(r)/r]/[3p(r)/2\rho(r)] $, is $3/2$ of the ratio of the absolute value of the gravitational potential energy, $|E_{g}| = Gm(r)/r$, and the internal energy per unit mass, $E_{i}=(3/2)p/\rho $, so that $v(r)=(3/2)|E_{g}|/E_{i}$~\cite{Horedt:2004,Blaga:2005}. In terms of these physical variables the deviation curvature tensor is given by
\begin{align}
  P_{1}^{1}(r)=\frac{1}{4}-\frac{3n}{2}\frac{\rho (r)}{\bar{\rho}(r)}\frac{\left|E_{g}\right|}{E_{i}}.
\end{align}
The condition of the Jacobi stability of the trajectories of the dynamical system describing a Newtonian polytropic star, $P_{1}^{1}<0$, can therefore be formulated as
\begin{align}
  \frac{E_{i}}{|E_{g}|} < 6 n \frac{\rho (r)}{\bar{\rho}(r)}.
\end{align}

\section{Lyapunov function stability analysis}
\label{sec:lya}

\subsection{Theory}

When discussing the mathematical background, we closely follow~\cite{Walter:1998,Khalil:2002}.

\begin{definition}[Lyapunov function]
Given a smooth dynamical system $\dot{x}=f(x)$, $x\in \mathbb{R}^{n}$, and an equilibrium point $x_{0}$, a continuous function $V:\mathbb{R}^{n}\rightarrow R$ in a neighborhood $U$ of $x_{0}$ is a Lyapunov function for the point $x_{0}$ if
\begin{enumerate}
\item $V$ is differentiable in $U\setminus \{ x_{0}\} $
\item $V(x)>V(x_{0})$
\item $\dot{V}(x) \leq 0$ for every $x\in U\setminus \left\{ x_{0}\right\}$.
\end{enumerate}
\end{definition}

The neighborhood $U$ is constrained by the third condition. When 3.~holds, the method provides information not only about the asymptotic stability of the equilibrium point, but also about its basin of attraction, which must contain the set $U$.

The existence of a Lyapunov function $V$ for an autonomous system of differential equations guarantees the stability of the point $x_{0}$. This qualitative result can be obtained without explicitly solving the equations, and this information cannot be obtained by using linear stability analysis or first order perturbation theory. The following theorem holds:

\begin{theorem}[Lyapunov stability]
Let $x_0$ be an equilibrium point of the system $\dot{x} = f(x)$, where $f : U \rightarrow \mathbb{R}^n$ is locally Lipschitz, and $U \subset \mathbb{R}^n$ is a domain that contains $x_0$. If $V$ is a Lyapunov function, then
\begin{enumerate}
\item if $\dot{V}(x) = \partial V/\partial x f$ is negative semi-definite,
then $x = x_0$ is a stable equilibrium point,
\item if $\dot{V}(x) = \partial V/\partial x f$ is negative definite,
then $x = x_0$ is an asymptotically stable equilibrium point.
\end{enumerate}
Moreover, if $\|x\| \rightarrow \infty$ implies that $V(x) \rightarrow \infty$ for all $x$, then $x_0$ is globally stable or globally asymptotically stable, respectively.
\end{theorem}

If the condition 3.~of the Lyapunov function definition holds strictly, the existence of a Lyapunov function has important consequences for the behavior of the time dependent perturbations of autonomous systems.

\subsection{Application to Emden-Fowler equation}

\begin{theorem}
 Let the dynamical system~(\ref{eqw}) and (\ref{syst}), with critical points~(\ref{critp1}) and~(\ref{critp3}), be given. The equilibrium state $X_0$ is asymptotically stable is $1 < n  < 3$, and the equilibrium state $X_n$ is asymptotically stable is $3 < n  < 5$.
\end{theorem}

\begin{proof}
One possible Lyapunov function $V(w,q)$ associated to the system given by Eqs.~(\ref{eqw}) and  (\ref{syst}) can be chosen following~\cite{Walter:1998}. Following the variable gradient method (this means setting $\nabla V = f \in \mathbb{R}^n$), we set
\begin{align}
  \nabla V(w,q) = \begin{pmatrix}
  -2(n-3)/(n-1)^{2}\,w + B^{n-1}w^{n} \\
  q
  \end{pmatrix}
\end{align}
such that the critical points correspond to $\nabla V = 0$ (since this corresponds to $f=0$). This yields the following Lyapunov function
\begin{align}
  V(w,q) =\frac{1}{2}q^{2}-\frac{(n-3)}{(n-1)^{2}}w^{2}+
  \frac{B^{n-1}}{n+1}w^{n+1}.
  \label{lyap}
\end{align}
By definition, the Lyapunov function must have a local minimum at the
critical points. To check this, we consider the Hessian of~(\ref{lyap}),
which is given by
\begin{align}
  H(V) = \begin{pmatrix}
  -2(n-3)/(n-1)^{2} + B^{n-1} n w^{n-1} & 0 \\
  0 & 1
  \end{pmatrix}
\end{align}
and has eigenvalues
\begin{align}
  \lambda_1 = -2(n-3)/(n-1)^{2} + B^{n-1} n w^{n-1},\qquad
  \lambda_2 = 1
\end{align}
At the point $X_0 = (0,0)$ for $n>1$ we have
\begin{align}
  \lambda_1 = -2(n-3)/(n-1)^{2},\qquad
  \lambda_2 = 1,
\end{align}
and hence there is a local minimum near $X_0$ if $ 1 < n < 3$.

At the point $X_n$, on the other hand, we find
\begin{align}
  \lambda_1 = 2(n-3)/(n-1),\qquad
  \lambda_2 = 1,
\end{align}
and hence there is a local minimum near $X_n$ if $n > 3$, and this function has no local minimum if the index satisfies $1 < n < 3$.

The Lyapunov function~(\ref{lyap}) satisfies
\begin{align}
  \frac{dV}{dt} = \frac{\partial V}{\partial w}\frac{dw}{dt}
  +\frac{\partial V}{\partial q}\frac{dq}{dt} =
  \frac{n-5}{n-1}q^{2},
\end{align}
and therefore we find that
\begin{align}
  \dot{V} < 0, \qquad 1< n < 5.
\end{align}
Hence, according to the Lyapunov theorem, the equilibrium state $X_0$ is asymptotically stable equilibrium points for $ 1 < n < 3$ while the equilibrium state $X_n$ is asymptotically stable equilibrium points for $ 3 < n < 5$. Global asymptotic stability results cannot be inferred from our Lyapunov function.
\end{proof}

\begin{remark}
Now we are faced with the difficulty of finding a Lyapunov function for other values of $n$. Recall that the Lyapunov theorem is in fact a quite restrictive statement, though very powerful. If stability properties cannot be established using a particular Lyapunov candidate function, it does not mean that the point is unstable.
\end{remark}

\section{Discussions and final remarks}
\label{sec:dis}

We considered the stability properties of the Emden-Fowler equation. For the stability analysis we used three methods: linear stability analysis, the so-called Jacobi stability analysis, or the KCC theory, and the Lyapunov function method. Our results show that  the study of the stability properties of the Emden-Fowler equation turns out to be a mathematically interesting problem.

\begin{table}[!ht]
\centering
\begin{tabular}{|c|c|c|c|}
\hline &&&\\[-2ex]
index $n$/method & Linear & Jacobi & Lyapunov function\\[1ex]
\hline\hline &&&\\[-2ex]
$1 < n < 3$ & unstable &  unstable & inconclusive\\[1ex]
\hline &&&\\[-2ex]
$3 < n < (11+8\sqrt{2})/7$ & stable & unstable & stable\\[1ex]
\hline&&&\\[-2ex]
$(11+8\sqrt{2})/7 < n < 5$ & stable & stable & stable\\[1ex]
\hline&&&\\[-2ex]
$5 < n$ & unstable & stable & inconclusive \\[1ex]
\hline
\end{tabular}
\caption{Stability properties of polytropic perfect fluid spheres with polytropic index $n$ as implied by linear perturbation theory, the Jacobi stability method (KCC theory), and the Lyapunov function method.}
\label{ta}
\end{table}

The study of the stability has been done by analyzing the behavior of the steady states $X_0$ and $X_n$. It is most interesting to note that these three method for some parameter region of the index $n$ give different stability properties. Our results can be summarized in Table~\ref{ta}.

Hence, we have found that a solution of the Emden-Fowler equation can be regarded as stable if and only if its polytropic index satisfies $(11+8\sqrt{2})/7 < n < 5$. For other parameter ranges the various methods used yield different results. It is however possible to give a good geometrical interpretation of the situation $3 < n < (11+8\sqrt{2})/7$ when the linear perturbation theory and the Lyapunov method both imply stability while the point is Jacobi unstable.

Let us recall~\cite{Sabau:2005aa} that the Jacobi stability of a dynamical system is regarded as the \textit{robustness} of the system to small perturbations of the \textit{whole} trajectory. This is a very convenient way of regarding the resistance of limit cycles to small perturbation of trajectories. It is interesting to remark that the stable nodal sink ($3 < n < (11+8\sqrt{2})/7$) in linear perturbation theory) is actually Jacobi unstable. In other words, even though the system's trajectories are attracted by the critical point $X_{n}$ one has to be aware of the fact that they are not stable to small perturbation of the whole trajectory.

We conclude by mentioning that our method can in principle be applied to all radially symmetric problems of the form $\Delta u + f(u) = 0$.

\subsection*{Acknowledgments}
We thank Steven Bishop for the useful discussions and valuable comments on the manuscript. The work of T.~H.~was supported by an RGC grant of the government of the Hong Kong SAR.


\providecommand{\bysame}{\leavevmode\hbox to3em{\hrulefill}\thinspace}
\providecommand{\MR}{\relax\ifhmode\unskip\space\fi MR }
\providecommand{\MRhref}[2]{%
  \href{http://www.ams.org/mathscinet-getitem?mr=#1}{#2}
}
\providecommand{\href}[2]{#2}

\end{document}